\documentclass[a4paper, conference]{IEEEtran}

\usepackage{cite}
\usepackage{amsmath,amssymb,amsfonts}
\usepackage{algorithmic}
\usepackage{graphicx}
\usepackage{textcomp}
\usepackage{xcolor}
\usepackage{paralist}
\usepackage{eurosym}
\usepackage{flushend}
\usepackage{url}
\usepackage[utf8]{inputenc}
\usepackage[colorlinks=true]{hyperref}
\usepackage{soul}
\usepackage{caption}
\usepackage[list=true]{subcaption}
\usepackage{booktabs}
\usepackage{tabu}
\usepackage{wrapfig}
\usepackage{arydshln}
\usepackage[normalem]{ulem}
\usepackage{multirow}
\usepackage{siunitx}
\usepackage{adjustbox}
\usepackage{array}
\usepackage{diagbox}
\usepackage{listings}

\def\rot{\rotatebox}

\newcommand\blfootnote[1]{%
  \begingroup
  \renewcommand\thefootnote{}\footnote{#1}%
  \addtocounter{footnote}{-1}%
  \endgroup
}

\begin{document}

\title{Physics-Informed Domain-Invariant Feature Learning with Autoencoder-Driven Gaussian Clustering for Robust Non-line-of-Sight Scenarios}

\author{\IEEEauthorblockN{Nisha L. Raichur\IEEEauthorrefmark{1}\IEEEauthorrefmark{2}, Lucas Heublein\IEEEauthorrefmark{1}, Dominik Seuß\IEEEauthorrefmark{1}\IEEEauthorrefmark{2}, Frank Deinzer\IEEEauthorrefmark{2}, Felix Ott\IEEEauthorrefmark{1}}
  \IEEEauthorblockA{\IEEEauthorrefmark{1}Fraunhofer Institute for Integrated Circuits IIS, 90411 Nürnberg, Germany}
  \IEEEauthorblockA{\IEEEauthorrefmark{2}Center for Artificial Intelligence, Technical University of Applied Sciences Würzburg-Schweinfurt, Germany}
  \IEEEauthorblockA{\{nisha.lakshmana.raichur, lucas.heublein, dominik.seuss, felix.ott\}@iis.fraunhofer.de, frank.deinzer@thws.de}
}

\maketitle

\begin{abstract}
Jamming and spoofing pose significant threats to wireless and satellite navigation by disrupting radio-frequency (RF) signals and compromising availability and integrity. Robust RF interference direction finding through angle-of-arrival (AoA) estimation is therefore essential for detecting and localizing anomalous signals. Although data-driven methods perform well under line-of-sight (LoS) conditions, their performance degrades in practical environments due to non-line-of-sight (NLoS) multipath propagation. In this work, we propose a hybrid learning framework that incorporates physics-informed constraints into deep neural networks to improve the robustness of AoA estimation. A neural network is trained to estimate the azimuth and elevation of incoming signals received by a four-element antenna array, while a physics-informed loss enforces consistency between the predicted angles and inter-antenna phase differences under a plane-wave model. We further introduce a latent-space classifier to distinguish LoS from NLoS samples. Since inter-antenna phase differences under LoS propagation exhibit domain-invariant structure across environments, the physics-based loss is applied only to LoS samples, promoting physically consistent and domain-invariant representations without over-constraining the model in NLoS scenarios. In addition, domain-incremental learning (DIL) across NLoS environments with varying scatterer distributions improves cross-domain generalization. Evaluations on real-world datasets show that the proposed method reduces AoA estimation error by up to $6^\circ$ in low-exemplar settings compared with DIL baselines.
\end{abstract}
\IEEEpeerreviewmaketitle

\section{Introduction}
\label{label_introduction}

The objective is to estimate the AoA of signals received by a multi-patch antenna and, from these measurements, localize the signal source relative to the array~\cite{kulakowski_vales_alonso}. An important application is the detection and direction finding of interference affecting global navigation satellite system (GNSS) signals. Classical AoA methods~\cite{schmidt1986multiple,roy2002esprit,barabell1983improving} estimate the impinging direction from array data using covariance-based array processing and signal/noise subspace separation, whereas ML methods~\cite{papageorgiou2021deep,heublein2025gnss,nguyen2022towards,yardibi2010source,heublein_wielenberg,dai2022deepaoanet} learn a direct mapping from multi-antenna measurements to angular coordinates from labeled data.

A major challenge for direction finding is multipath propagation, which occurs when signals are reflected, scattered, or partially blocked by surrounding objects such as buildings, terrain, vehicles, or vegetation, so the receiver observes multiple paths instead of a single LoS component~\cite{zhao_xu_ding}. This distorts the phase and amplitude relationships across the array and degrades AoA estimation, particularly across environments such as cities, forests, and open fields, where propagation conditions vary substantially~\cite{ahmed_arablouei}. Robust systems must therefore identify domain shifts present across environments and adapt accordingly to novel scenarios.

DIL addresses this challenge by sequentially adapting a model to new domains with shifted data distributions while preserving previously acquired knowledge~\cite{shen2025domain,raichur_heublein}. In environments with different multipath characteristics, DIL updates the model using limited data from each new domain together with selected exemplars from previous domains, enabling adaptation without catastrophic forgetting~\cite{raichur20255g,raichur2026adaptive}. Physics-informed neural networks~\cite{benelmekki2026physics,javid2025physics,raissi2019physics,rasht2022physics,liu2024phase,zhang2026physics} further improve AoA estimation by embedding physical laws into the training objective, enforcing consistency between predicted angles and the array geometry or inter-antenna phase relationships under challenging propagation conditions. By combining continual adaptation with physics-based regularization, these approaches improve robustness for AoA estimation under multipath effects.

\textbf{Contributions.} In the following, we summarize our main contributions: (1) We propose a hybrid method for GNSS interference localization and AoA estimation that combines deep regression with physics-informed regularization. (2) We introduce an unsupervised LoS/NLoS identification pipeline based on statistical IQ features, autoencoder (AE)-based latent representations, and Gaussian mixture model (GMM) clustering. (3) We develop a selective physics-informed loss based on the plane-wave model and inter-antenna phase differences, applied only to LoS-dominant samples through a probabilistic gating mechanism to promote physically consistent and domain-invariant feature learning without over-constraining the model under NLoS conditions. (4) We show that SHAP-guided feature reduction preserves the most discriminative information for LoS/NLoS separation while substantially reducing input dimensionality and computational complexity. (5) We record a novel indoor dataset, formulate adaptation across environments with increasing wall-induced multipath as a DIL problem, and demonstrate a low-exemplar strategy for efficient sequential adaptation while retaining previously acquired knowledge.
\section{Related Work}
\label{label_related_work}

Traditional AoA estimation is typically based on the eigenspace structure of the received-signal covariance matrix. Classical subspace-based methods, including MUSIC~\cite{schmidt1986multiple}, ESPRIT~\cite{roy2002esprit}, and Root-MUSIC~\cite{barabell1983improving}, can provide high-resolution estimates under favorable conditions. However, their performance generally depends on accurate array modeling and calibration, and it can degrade markedly in low-SNR or multipath-dominated environments. Recently, ML-based approaches have increasingly been employed to address these limitations by directly mapping signal features or covariance matrices to angular estimates~\cite{papageorgiou2021deep,dai2022deepaoanet}. Such methods have shown strong performance in anti-jamming applications~\cite{heublein2025gnss,nguyen2022towards} and in data-driven, non-parametric sensing frameworks~\cite{yardibi2010source}.

To improve the robustness of regression-based models across multiple target scenarios, especially under overlapping data distributions, DIL strategies have been introduced. In particular, regularization-based methods such as elastic weight consolidation (EWC)~\cite{kirkpatrick_rascanu} and memory aware synapses (MAS)~\cite{aljundi2018memory} mitigate catastrophic forgetting by identifying important parameters and constraining their updates during training~\cite{raichur2026adaptive}. Distillation~\cite{li_hoiem} preserves prior-task knowledge by minimizing the discrepancy between the outputs of the current and previously trained models on new-task data. However, it does not explicitly incorporate electromagnetic propagation principles, which may limit generalization under mismatched deployment conditions.

To bridge empirical learning and physical consistency, physics-informed networks have emerged as a powerful framework~\cite{raissi2019physics}. By embedding governing equations or physical constraints into the training objective, physics-informed networks enforce consistency with known laws, including wave propagation and field reconstruction~\cite{rasht2022physics, zhang2026physics}. In the RF domain, they have been applied to wave prediction~\cite{liu2024phase}, wireless channel estimation~\cite{javid2025physics}, and AoA estimation, where Bayesian physics-informed constraints have achieved sub-degree RFID accuracy~\cite{benelmekki2026physics}. Building on this idea, our approach leverages the plane-wave assumption to regularize the latent space and employs a gating mechanism to disable this constraint under NLoS conditions, where the assumption is violated. To address temporal and spatial domain shifts, we further adopt DIL to continuously adapt the model while mitigating catastrophic forgetting~\cite{raichur20255g}. Recent advances in 5G indoor localization have explored similarity-aware sampling and adaptive selection strategies to address dynamic multipath environments~\cite{raichur2026adaptive}. Transfer learning and AE-based methods have been used to adapt AoA models to complex indoor environments~\cite{shen2025domain}. Unlike these purely statistical approaches, our method learns domain-invariant features by enforcing physical consistency on LoS components, which remain comparatively stable across environments. By grounding the learning process in physical principles, the proposed framework achieves stronger generalization in low-exemplar scenarios, consistent with recent work on similarity-aware sampling and DIL localization under changing environments.
\section{Methodology}
\label{label_method}

We begin by introducing the notation (Section~\ref{label_method_notation}) and an overview of our proposed methodology (Section~\ref{label_method_overview}). We then describe the LoS/NLoS classification framework (Section~\ref{label_method_los_nlos}) and conclude by detailing the physics-informed loss used to learn domain-invariant features (Section~\ref{label_method_physics_informed}).

\subsection{Notation}
\label{label_method_notation}

The primary objective is to estimate the relative position and AoA of a GNSS interference source w.r.t~the receiver antenna array. We estimate the relative position in 3D-space, $\hat{\mathbf{p}} = (\Delta\hat{x}, \Delta\hat{y}, \Delta\hat{z}) \in \mathbb{R}^3$, and the angular components azimuth $\alpha$ and elevation $\beta$. We encode each angle as a unit vector on a circle rather than regressing the angle directly. Denoting the azimuth and elevation in radians as $\theta_{\alpha}$ and $\theta_{\beta}$, we define the AoA targets as $\mathbf{\hat{e}}_{\alpha} = (\sin\theta_{\alpha}, \cos\theta_{\alpha}) \in \mathbb{R}^2$, and $\mathbf{\hat{e}}_{\beta} = (\sin\theta_{\beta}, \cos\theta_{\beta}) \in \mathbb{R}^2$, The ground truth values are denoted as $\mathbf{p} = (\Delta x, \Delta y, \Delta z) \in \mathbb{R}^3$, $\mathbf{e}_{\alpha}$ and $\mathbf{e}_{\beta}$. The input to the model consists of complex in-phase and quadrature (IQ) samples collected from a $2 \times 2$ antenna array with $M=4$ elements. For each sample, the multi-antenna observation is represented as $\mathbf{x}[n] = \big[x_1[n], x_2[n], x_3[n], x_4[n]\big]^T \in \mathbb{C}^4$, where $x_i[n]$ denotes the complex IQ sample at the $i$-th antenna element and discrete time index $n$. The corresponding phase of each antenna signal is denoted by $\phi_i[n] = \angle x_i[n]$. Since absolute phase is ambiguous, relative phase differences between antenna elements are considered. Taking the first antenna as a reference, the relative phase differences are denoted as $\Delta \boldsymbol{\phi}[n] = \big[\Delta \phi_2[n], \Delta \phi_3[n], \Delta \phi_4[n]\big]^T$, $\forall i \in \{2, 3, 4\}$, where $\Delta \phi_i[n]=\phi_i[n]-\phi_1[n]$ represents the phase difference between the $i$-th antenna and $\phi_1[n]$.

In incremental learning scenarios, a sequence of tasks is learned, each corresponding to a distinct propagation environment characterized by different wall configurations and scatterer distributions. This can be defined as a sequence of $k$ tasks $\mathcal{T} = \big[(D^1, Y^1), (D^2, Y^2), \ldots, (D^k, Y^k)\big]$, where for each task $t$, the dataset is given by
\begin{equation}
\resizebox{0.91\linewidth}{!}{$
    (D^t, Y^t) = \big\{ (\mathbf{x}_1^t[n], \mathbf{y}_1^t), (\mathbf{x}_2^t[n], \mathbf{y}_2^t), \ldots, (\mathbf{x}_{k}^t[n], \mathbf{y}_{\nu^{t}}^t) \big\},
$}
\end{equation}
where $\nu^{t}$ represents the number of data points in each task $t$.

\subsection{Method Overview}
\label{label_method_overview}

\begin{figure*}[!t]
    \centering
    \includegraphics[width=1.0\linewidth]{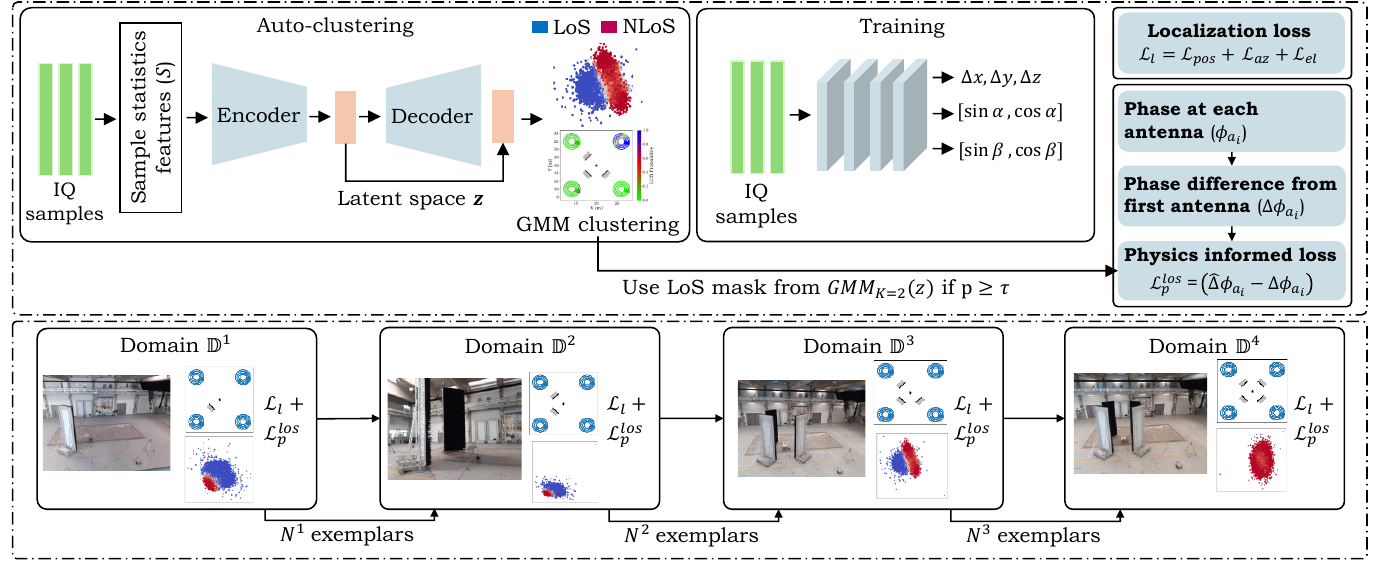}
    \vspace{-0.5cm}
    \caption{\textbf{Method overview.} IQ samples are encoded into a latent space and clustered to identify LoS/NLoS. A LoS mask activates the physics-informed loss during AoA training, while DIL adapts the model across environments using limited exemplars.}
    \label{figure_pipeline_overview}
\end{figure*}

Figure~\ref{figure_pipeline_overview} provides a comprehensive overview of the proposed method. Initially, 22 statistical features are extracted from the raw IQ samples. An AE is then trained to learn a compact latent representation, which subsequently serves as the input to GMM clustering. The resulting latent features are used for LoS/NLoS clustering, upon which all loss functions are defined and optimized. For each task $t$, the model processes the IQ measurements $D^t$ along with the corresponding reference positions and AoA labels ${Y}^t$. The primary objective of our approach is to estimate $\hat{\mathbf{p}} \in \mathbb{R}^3$, $\hat{\alpha} \in \mathbb{R}^2$  and $\hat{\beta} \in \mathbb{R}^2$ of an incoming interference signal. To achieve this, we employ the XceptionTime model~\cite{rahimian_zabihi} from the \texttt{tsai} library~\cite{tsai}. This constitutes the primary loss of the system, denoted as the \textit{localization loss} $\mathcal{L}_l$. In addition, the proposed framework is designed to learn a sequence of domains corresponding to distinct environmental scenarios. In this study, each domain is defined by a different number of walls, which induce varying levels of multipath propagation. The sequential acquisition of these domains is formulated as a DIL problem. Each scenario is represented by a four-circle grid, with data initially collected over all circles. The model is first trained on the single-wall scenario and subsequently fine-tuned on scenarios with two, three, and four walls. 

Formally, let $\mathcal{D}_t$ denote the dataset corresponding to the $t$-th domain (number of walls). The model is first trained on $\mathcal{D}_1$ to minimize the localization loss:
\begin{equation}
    \mathcal{L}_{\text{l}} = \text{MSE}(\hat{\mathbf{p}}, \mathbf{p}) + \text{MSE}(\mathbf{\hat{e}}_{\alpha}, \mathbf{e}_{\alpha}) + \text{MSE}(\mathbf{\hat{e}}_{\beta}, \mathbf{e}_{\beta}),
\end{equation}
where $\mathbf{p}$, $\mathbf{e}_{\alpha}$, and $\mathbf{e}_{\beta}$ denote the direction finding of the interference source, $\hat{\cdot}$ indicates the model predictions, and MSE defines the mean squared error between ground truth and predictions. To enable domain-invariant feature learning, we introduce a physics-informed loss, $\mathcal{L}_{\text{p}}$, described in Section~\ref{label_method_physics_informed}, which encourages the model to capture features that are robust across different domains. The final objective function combines the localization loss and the physics-informed loss by $\mathcal{L}_{\text{total}} = \mathcal{L}_{\text{l}} + \lambda \mathcal{L}_{\text{p}}$, where $\lambda$ is a hyperparameter controlling the relative importance of the physics-informed loss. The model is then fine-tuned incrementally on $\mathcal{D}_2$, $\mathcal{D}_3$, and $\mathcal{D}_4$, allowing the model to sequentially adapt to novel environmental configurations while preserving previously learned knowledge.

\subsection{LoS \& NLoS Classification}
\label{label_method_los_nlos}

To distinguish between LoS and NLoS propagation conditions, we employ an unsupervised feature-based classification framework inspired by prior work on AoA feature engineering~\cite{heublein2025gnss}. For each antenna element, a set of 22 commonly used statistical features is extracted following the framework of Wu et al.~\cite{wu_zhou_shen}. These features capture complementary properties of the received signal and are organized into six categories: temporal, spectral, energy-based, envelope-based, phase-difference-based, and IQ-component-based features. The resulting features are concatenated across the $M=4$ antenna elements to form an input vector $\mathbf{sf} \in \mathbb{R}^{88}$, with each antenna contributing $n_{f}=22$ features. This vector is then used to learn a compact latent representation through an AE.

\paragraph{Latent Representation Learning} We employ a fully connected AE to compress the high-dimensional feature vector into a low-dimensional latent space. The encoder maps $\mathbf{sf}_{n_{f}} \in \mathbb{R}^{n_{f}\times4}$ to a latent vector $\mathbf{z} \in \mathbb{R}^{d}$, where $n_f$ is the number of input features, and $d = 4$ is the latent dimensionality, while the decoder reconstructs the input features. The model is trained to minimize the reconstruction error, encouraging the latent space to capture the most informative structure of the data:
\begin{equation}
\label{eq_autoencoder}
    \mathbf{z} = \text{f}_{\text{enc}}(\mathbf{sf}), \quad \hat{\mathbf{f}} = \text{f}_{\text{dec}}(\mathbf{z}).
\end{equation}

\paragraph{Unsupervised LoS \& NLoS Classification} The learned latent representation $\mathbf{z}$ is subsequently used for unsupervised classification. Specifically, we fit a GMM~\cite{reynolds2009gaussian} with two components and full covariance matrices to the latent space $\mathbf{z}$ by the mapping
\begin{equation}
\label{eq_gmm}
    \mathbf{p} = \mathrm{GMM}_{K=2}(\mathbf{z}),
\end{equation}
where $\mathrm{GMM}_{K=2} {:} \,\, \mathbb{R}^{d} \rightarrow [0,1]^2$ outputs the posterior probabilities of the LoS and NLoS classes 0 and 1, respectively, using expectation-maximization.

\paragraph{Feature Importance Analysis} To interpret the contribution of individual features to the LoS and NLoS classification, we employ SHAP (SHapley Additive exPlanations)~\cite{lundberg_lee} with a kernel explainer. The kernel explainer operates on the end-to-end unsupervised pipeline, mapping input features  $\mathbf{sf}$ to the predicted LoS probability $\mathbf{p}$ by
\begin{equation}
\label{eq_kernelshap}
    \boldsymbol{\psi} = \mathrm{KernelExplainer}\Big(\mathrm{GMM}_{K=2}\big(f_{\text{enc}}(\mathbf{sf_{n_{f}}})\big), \mathbf{sf_{n_{f}}}\Big),
\end{equation}
which attributes the contribution of $\mathbf{sf}$ to $\mathbf{p}$. Using a representative subset of samples, we compute SHAP values $\boldsymbol{\psi} \in \mathbb{R}^{88}$ to quantify the contribution of each feature $\boldsymbol{\psi}$  to the classification outcome. The analysis reveals that a small subset of features $\mathcal{S}=4$ consistently dominates the prediction, namely phase difference standard deviation, kurtosis of the in-phase (I) component, mean of the imaginary (Q) component and mean of the real (I) component.

\paragraph{Feature Reduction} Motivated by the SHAP analysis, we construct a reduced feature set $\mathcal{S}$ by retaining only the most influential features from each antenna, thereby yielding a substantially lower-dimensional input representation. The AE is then retrained using this reduced feature set, and the same latent-space GMM-based classification procedure is applied. Empirical results indicate that the reduced representation achieves performance comparable to that of the full feature set, suggesting that the selected features preserve the most discriminative information for LoS/NLoS separation. This reduction not only simplifies the model but also lowers computational complexity while maintaining classification performance.

\subsection{Physics-Informed Domain-Invariant Feature Learning}
\label{label_method_physics_informed}

To improve the model’s ability to generalize across heterogeneous environmental domains, we introduce a physics-informed loss that exploits the known structure of the LoS components of interference sources. Because LoS propagation is comparatively less affected by multipath reflections from walls and other obstacles, these components provide a reliable basis for learning domain-invariant features. Let $\mathbf{AoA}_{\text{pred}} = (\hat{\alpha}, \hat{\beta})$ and $\mathbf{AoA}_{\text{label}} = (\alpha, \beta)$ denote the predicted and ground truth angles of arrival for a batch of samples. We first convert these angles to unit vectors:
\begin{equation}
    \mathbf{u} = \big[ \cos\beta\cos\alpha, \cos\beta\sin\alpha, \sin\beta \big]^T.
\end{equation}
The phases across array elements are then computed as $\phi = \frac{2 \pi}{\lambda} (\mathbf{u} \mathbf{R}^\top)$, where $\lambda = 0.1903\,\text{m}$  is the wavelength and $\mathbf{R}$ contains the  $2{\times}2$ array element positions represented as 
\begin{equation}
\resizebox{0.6\linewidth}{!}{$
    \mathbf{R} = 
    \begin{bmatrix}
    -d/2 & -d/2 & 0 \\
    -d/2 & d/2 & 0 \\
    d/2 & d/2 & 0 \\
    d/2 & -d/2 & 0
    \end{bmatrix}, \quad d = 0.09\,\text{m}.
$}
\end{equation}
The phase differences are computed as $\Delta \phi_i[n] = \phi_i[n] - \phi_1[n]$, for $i \in \{2,3,4\}$, where each term represents the phase difference between antenna $i$ and the reference antenna. To further emphasize LoS-dominant samples, we employ a probabilistic LoS mask derived from $\text{f}_{\text{enc}}$ (see Equation~\ref{eq_autoencoder}) and the GMM (see Equation~\ref{eq_gmm}). Only samples with an LoS probability exceeding a threshold $\tau$ are retained. Denoting this mask by $M_{\text{LoS}}$, the physics-informed loss is defined as
\begin{equation}
    \mathcal{L}_{\text{p}} = 
    \frac{1}{|M_{\text{LoS}}|} \sum_{j \in M_{\text{LoS}}} 
    \Big[\arctan_2 \big( \sin(\delta_j), \cos(\delta_j) \big) \Big]^2,
\end{equation}
where the mean squared error is computed between the predicted and ground-truth phase differences, with \(\delta_j = \hat{\Delta}\phi_j - \Delta\phi_j\), for the LoS components. This loss encourages accurate prediction of LoS phase differences and, in turn, promotes the learning of features that remain invariant across domains, since LoS propagation is comparatively stable even as wall configurations change.
\section{Experiments}
\label{label_experiments}

\paragraph{Dataset} To enable a comprehensive evaluation of the proposed method, we conduct a large-scale measurement campaign at the Fraunhofer IIS test center, which offers an industrial environment with realistic multipath propagation characteristics. The experimental setup combines an Ettus USRP X440 software-defined radio (SDR) platform with a 3D dynamic positioning system, while a stationary jammer located at the center of the scenario serves as the interference source. Using the SDR platform, measurements are recorded at a center frequency of $1.57542\,\text{GHz}$ with a bandwidth of $40.96\,\text{MHz}$. Data are acquired in $3\,\text{ms}$ snapshots at a rate of five snapshots per second and streamed via dual $100\,\text{Gb}$ Ethernet to a high-performance system. A square antenna array comprising four cellular antennas with a side length of $9\,\text{cm}$ is mounted on the 3D platform. The receiver follows predefined circular trajectories within a 3D measurement volume at a maximum velocity of $0.3\,\text{m/s}$. For each wall configuration, four spatial grid layouts are recorded with the jammer positioned at the center. In every scenario, absorber/reflecting walls are placed between the interference source and the receiver, as shown in Figure~\ref{figure_pipeline_overview}. For each grid, five concentric circular trajectories with radii between $0.8\,\text{m}$ and $2\,\text{m}$ are executed at four heights ranging from $3.3\,\text{m}$ to $4.8\,\text{m}$, resulting in dense 3D sampling of the environment. Ground-truth annotations are provided by the positioning system and include the relative spatial coordinates and direction angles between the receiver and the interference source. 

\paragraph{DIL Setup} To examine the impact of varying multipath conditions, we define multiple domains characterized by distinct wall-induced propagation effects. While the overall experimental setup remains unchanged, comprising a static jammer and a mobile receiver, the domains differ in their multipath structure due to the number of walls placed between transmitter and receiver. We consider a DIL setting in which the model is sequentially exposed to data from multiple domains. In this work, four domains are defined (i.e., with one, two, three, or four walls), refer to Figure~\ref{figure_pipeline_overview}. For each domain, measurements are collected over four spatial grid configurations, ensuring variability in geometric relationships while preserving the underlying domain characteristics.

\begin{table*}[!t]
\begin{center}
\setlength{\tabcolsep}{3.7pt}
    \caption{Evaluation after sequential training on 1 $\rightarrow$ 2 $\rightarrow$ 3 $\rightarrow$ 4 walls. Last three columns: average MSE across all domains.}
    \label{table_dil_results}
    \vspace{-0.1cm}
    \scriptsize \begin{tabular}{ p{0.25cm} p{0.25cm} p{0.25cm} p{0.25cm} p{0.25cm} p{0.25cm} p{0.25cm} p{0.25cm} p{0.25cm} p{0.25cm} p{0.25cm} p{0.25cm} p{0.25cm} p{0.25cm} p{0.25cm} p{0.25cm} p{0.25cm} p{0.25cm} p{0.25cm} }
    \toprule
    \multicolumn{1}{c}{\textbf{Buffer}} &
    \multicolumn{1}{c}{\textbf{Method/}} &
    \multicolumn{1}{c}{\textbf{Features}}
     & \multicolumn{3}{c}{\textbf{1 wall}}
     & \multicolumn{3}{c}{\textbf{2 walls}}
     & \multicolumn{3}{c}{\textbf{3 walls}}
     & \multicolumn{3}{c}{\textbf{4 walls}}
     & \multicolumn{3}{c}{\textbf{Average}} \\
     \multicolumn{1}{c}{\textbf{Size}} &
     \multicolumn{1}{c}{\textbf{Loss}} &
     \multicolumn{1}{c}{$n_f$} 
     & \multicolumn{1}{c}{$\mathcal{E}_{p}$} 
     & \multicolumn{1}{c}{$\mathcal{E}_{a}$} 
     & \multicolumn{1}{c}{$\mathcal{E}_{e}$}
     & \multicolumn{1}{c}{$\mathcal{E}_{p}$} 
     & \multicolumn{1}{c}{$\mathcal{E}_{a}$} 
     & \multicolumn{1}{c}{$\mathcal{E}_{e}$}
     & \multicolumn{1}{c}{$\mathcal{E}_{p}$}
     & \multicolumn{1}{c}{$\mathcal{E}_{a}$} 
     & \multicolumn{1}{c}{$\mathcal{E}_{e}$}
     & \multicolumn{1}{c}{$\mathcal{E}_{p}$}
     & \multicolumn{1}{c}{$\mathcal{E}_{a}$} 
     & \multicolumn{1}{c}{$\mathcal{E}_{e}$} 
     & \multicolumn{1}{c}{\textbf{$\overline{\mathcal{E}_{p}}$}}
     & \multicolumn{1}{c}{\textbf{$\overline{\mathcal{E}_{a}}$}} 
     & \multicolumn{1}{c}{\textbf{$\overline{\mathcal{E}_{e}}$}}
     \\
    \midrule
     \multicolumn{1}{c}{All} & 
     \multicolumn{1}{l}{Regular} & 
     \multicolumn{1}{c}{IQ} & 
    \multicolumn{1}{r}{3.75} &
    \multicolumn{1}{r}{74.95} &
    \multicolumn{1}{r}{4.39} &
    \multicolumn{1}{r}{3.64} &
    \multicolumn{1}{r}{67.77} &
    \multicolumn{1}{r}{4.32} &
    \multicolumn{1}{r}{3.20} &
    \multicolumn{1}{r}{53.94} &
    \multicolumn{1}{r}{3.74} &
    \multicolumn{1}{r}{0.98} &
    \multicolumn{1}{r}{6.70} & 
    \multicolumn{1}{r}{3.31} &
    \multicolumn{1}{r}{2.89} &
    \multicolumn{1}{r}{50.84} & 
    \multicolumn{1}{r}{3.94} \\
    \midrule
     \multirow{6}{*}{\rot{90}{1,000}} &  
     \multicolumn{1}{l}{Regular} & 
     \multicolumn{1}{c}{IQ} & 
    \multicolumn{1}{r}{1.66} &
    \multicolumn{1}{r}{20.04} &
    \multicolumn{1}{r}{3.47} &
    \multicolumn{1}{r}{1.52} &
    \multicolumn{1}{r}{17.62} &
    \multicolumn{1}{r}{3.46} &
    \multicolumn{1}{r}{1.90} &
    \multicolumn{1}{r}{21.56} &
    \multicolumn{1}{r}{3.34} &
    \multicolumn{1}{r}{0.90} &
    \multicolumn{1}{r}{6.14} & 
    \multicolumn{1}{r}{3.18} & 
    \multicolumn{1}{r}{1.50} &
    \multicolumn{1}{r}{16.34} & 
    \multicolumn{1}{r}{3.36}\\
    \multicolumn{1}{r}{} &
    \multicolumn{1}{l}{Distillation~\cite{li_hoiem}} & 
    \multicolumn{1}{c}{IQ} & 
    \multicolumn{1}{r}{1.46} &
     \multicolumn{1}{r}{16.44} &
     \multicolumn{1}{r}{3.20} & 
     \multicolumn{1}{r}{1.48}& 
     \multicolumn{1}{r}{16.57}& 
    \multicolumn{1}{r}{3.14} & 
    \multicolumn{1}{r}{1.97} &
    \multicolumn{1}{r}{21.49} &
    \multicolumn{1}{r}{3.34} &
    \multicolumn{1}{r}{0.90} &
    \multicolumn{1}{r}{6.18} &
    \multicolumn{1}{r}{3.18} &
    \multicolumn{1}{r}{1.45} &
    \multicolumn{1}{r}{15.17} &
    \multicolumn{1}{r}{3.22}\\
    \multicolumn{1}{r}{} &
    \multicolumn{1}{l}{EWC~\cite{kirkpatrick_rascanu}} & 
    \multicolumn{1}{c}{IQ} &
    \multicolumn{1}{c}{1.48} & 
     \multicolumn{1}{r}{17.30} &
     \multicolumn{1}{r}{3.34} & 
     \multicolumn{1}{r}{1.48}& 
     \multicolumn{1}{r}{16.68}& 
     \multicolumn{1}{r}{3.38}& 
    \multicolumn{1}{r}{1.85} & 
    \multicolumn{1}{r}{18.32} &
    \multicolumn{1}{r}{3.70} &
    \multicolumn{1}{r}{0.91} &
    \multicolumn{1}{r}{6.24} &
    \multicolumn{1}{r}{3.17} &
    \multicolumn{1}{r}{1.43} &
    \multicolumn{1}{r}{14.64} &
    \multicolumn{1}{r}{3.40}\\
    \multicolumn{1}{r}{} &
    \multicolumn{1}{l}{MAS~\cite{aljundi2018memory}} & 
    \multicolumn{1}{c}{IQ} & 
     \multicolumn{1}{r}{1.55} &
     \multicolumn{1}{r}{19.32} & 
     \multicolumn{1}{r}{4.06}& 
     \multicolumn{1}{r}{1.54}& 
     \multicolumn{1}{r}{17.97}& 
    \multicolumn{1}{r}{3.74} & 
    \multicolumn{1}{r}{2.06} &
    \multicolumn{1}{r}{21.11} &
    \multicolumn{1}{r}{3.52} &
    \multicolumn{1}{r}{1.11} &
    \multicolumn{1}{r}{7.88} &
    \multicolumn{1}{r}{3.35} &
    \multicolumn{1}{r}{1.57} &
    \multicolumn{1}{r}{16.57} &
    \multicolumn{1}{r}{3.67}\\    
    \multicolumn{1}{r}{} &
    \multicolumn{1}{l}{Physics-Informed (ours)} & 
    \multicolumn{1}{c}{22} & 
     \multicolumn{1}{r}{1.44} &
     \multicolumn{1}{r}{16.16} & 
     \multicolumn{1}{r}{3.14}& 
     \multicolumn{1}{r}{1.46}& 
     \multicolumn{1}{r}{15.46}& 
    \multicolumn{1}{r}{3.21} & 
    \multicolumn{1}{r}{1.79} &
    \multicolumn{1}{r}{14.89} &
    \multicolumn{1}{r}{3.26} &
    \multicolumn{1}{r}{0.89} &
    \multicolumn{1}{r}{6.05} &
    \multicolumn{1}{r}{3.16} &
    \multicolumn{1}{r}{\textbf{1.40}} &
    \multicolumn{1}{r}{\textbf{13.14}} &
    \multicolumn{1}{r}{\textbf{3.19}}\\
    \multicolumn{1}{r}{} &
    \multicolumn{1}{l}{Physics-Informed (ours)} & 
    \multicolumn{1}{c}{4} & 
     \multicolumn{1}{r}{1.47} &
     \multicolumn{1}{r}{16.08} & 
     \multicolumn{1}{r}{3.26}& 
     \multicolumn{1}{r}{1.45}& 
     \multicolumn{1}{r}{15.43}& 
    \multicolumn{1}{r}{3.21} & 
    \multicolumn{1}{r}{1.80} &
    \multicolumn{1}{r}{16.44} &
    \multicolumn{1}{r}{3.37} &
    \multicolumn{1}{r}{0.88} &
    \multicolumn{1}{r}{6.04} &
    \multicolumn{1}{r}{3.17} &
    \multicolumn{1}{r}{1.40} &
    \multicolumn{1}{r}{13.50} &
    \multicolumn{1}{r}{3.25}\\
    \midrule
    \multirow{6}{*}{\rot{90}{500}} &
   \multicolumn{1}{l}{Regular} & 
   \multicolumn{1}{c}{IQ} & 
    \multicolumn{1}{r}{1.68} &
    \multicolumn{1}{r}{22.01} &
    \multicolumn{1}{r}{3.43} &
    \multicolumn{1}{r}{1.67} &
    \multicolumn{1}{r}{22.29} &
    \multicolumn{1}{r}{3.46} &
    \multicolumn{1}{r}{1.99} &
    \multicolumn{1}{r}{25.25} &
    \multicolumn{1}{r}{3.34} &
    \multicolumn{1}{r}{0.88} &
    \multicolumn{1}{r}{6.14} & 
    \multicolumn{1}{r}{3.18} & 
    \multicolumn{1}{r}{1.56} &
    \multicolumn{1}{r}{19.12} & 
    \multicolumn{1}{r}{3.35}\\
    \multicolumn{1}{r}{} &
    \multicolumn{1}{l}{Distillation~\cite{li_hoiem}} & 
    \multicolumn{1}{c}{IQ} & 
     \multicolumn{1}{r}{1.99} &
     \multicolumn{1}{r}{24.91} & 
     \multicolumn{1}{r}{3.56} & 
     \multicolumn{1}{r}{1.82} & 
     \multicolumn{1}{r}{23.39} & 
    \multicolumn{1}{r}{3.56} & 
    \multicolumn{1}{r}{1.95} &
    \multicolumn{1}{r}{18.20} &
    \multicolumn{1}{r}{3.40} &
    \multicolumn{1}{r}{1.07} &
    \multicolumn{1}{r}{7.81} &
    \multicolumn{1}{r}{3.25} &
    \multicolumn{1}{r}{1.71} &
    \multicolumn{1}{r}{18.58} &
    \multicolumn{1}{r}{3.44} \\
    \multicolumn{1}{r}{} &
    \multicolumn{1}{l}{EWC~\cite{kirkpatrick_rascanu}} & 
    \multicolumn{1}{c}{IQ} &
    \multicolumn{1}{r}{1.69} & 
     \multicolumn{1}{r}{22.28} &
     \multicolumn{1}{r}{3.44} & 
     \multicolumn{1}{r}{1.66} & 
     \multicolumn{1}{r}{21.53}& 
     \multicolumn{1}{r}{3.40}& 
    \multicolumn{1}{r}{1.97} & 
    \multicolumn{1}{r}{21.86} &
    \multicolumn{1}{r}{3.37} &
    \multicolumn{1}{r}{0.89} &
    \multicolumn{1}{r}{6.10} &
    \multicolumn{1}{r}{3.15} &
    \multicolumn{1}{r}{1.55} &
    \multicolumn{1}{r}{17.94} &
    \multicolumn{1}{r}{3.34}\\
    \multicolumn{1}{r}{} &
    \multicolumn{1}{l}{MAS~\cite{aljundi2018memory}} & 
    \multicolumn{1}{c}{IQ} &
    \multicolumn{1}{r}{1.91} & 
     \multicolumn{1}{r}{23.63} &
     \multicolumn{1}{r}{4.05} & 
     \multicolumn{1}{r}{1.85}& 
     \multicolumn{1}{r}{21.19}& 
     \multicolumn{1}{r}{3.98}& 
    \multicolumn{1}{r}{2.27} & 
    \multicolumn{1}{r}{23.14} &
    \multicolumn{1}{r}{3.75} &
    \multicolumn{1}{r}{1.36} &
    \multicolumn{1}{r}{10.01} &
    \multicolumn{1}{r}{3.41} &
    \multicolumn{1}{r}{1.85} &
    \multicolumn{1}{r}{19.50} &
    \multicolumn{1}{r}{3.80} 
    \\
    \multicolumn{1}{r}{} &
     \multicolumn{1}{l}{Physics-Informed (ours)} & 
     \multicolumn{1}{c}{4} & 
     \multicolumn{1}{r}{1.58} &
     \multicolumn{1}{r}{18.55} & 
     \multicolumn{1}{r}{3.36}& 
     \multicolumn{1}{r}{1.61}& 
     \multicolumn{1}{r}{19.13}& 
    \multicolumn{1}{r}{3.39} & 
    \multicolumn{1}{r}{1.87} &
    \multicolumn{1}{r}{16.18} &
    \multicolumn{1}{r}{3.23} &
    \multicolumn{1}{r}{0.88} &
    \multicolumn{1}{r}{5.95} &
    \multicolumn{1}{r}{3.16} &
    \multicolumn{1}{r}{\textbf{1.49}} &
    \multicolumn{1}{r}{\textbf{14.95}} &
    \multicolumn{1}{r}{\textbf{3.29}}\\
    \midrule
   \multirow{6}{*}{\rot{90}{100}} & 
    \multicolumn{1}{l}{Regular} & 
    \multicolumn{1}{c}{IQ} & 
    \multicolumn{1}{r}{2.20} &
    \multicolumn{1}{r}{32.71} &
    \multicolumn{1}{r}{3.59} &
    \multicolumn{1}{r}{2.21} &
    \multicolumn{1}{r}{31.06} &
    \multicolumn{1}{r}{3.50} &
    \multicolumn{1}{r}{2.15} &
    \multicolumn{1}{r}{34.30} &
    \multicolumn{1}{r}{3.33} &
    \multicolumn{1}{r}{0.90} &
    \multicolumn{1}{r}{6.02} & 
    \multicolumn{1}{r}{3.16} & 
    \multicolumn{1}{r}{1.93} &
    \multicolumn{1}{r}{26.07} & 
    \multicolumn{1}{r}{3.40}\\
    \multicolumn{1}{r}{} &
    \multicolumn{1}{l}{Distillation~\cite{li_hoiem}} & 
    \multicolumn{1}{c}{IQ} & 
     \multicolumn{1}{r}{2.25} &
     \multicolumn{1}{r}{28.33} & 
     \multicolumn{1}{r}{4.02}& 
     \multicolumn{1}{r}{2.35}& 
     \multicolumn{1}{r}{33.30}& 
    \multicolumn{1}{r}{4.24} & 
    \multicolumn{1}{r}{2.43} &
    \multicolumn{1}{r}{36.27} &
    \multicolumn{1}{r}{4.11} &
    \multicolumn{1}{r}{1.15} &
    \multicolumn{1}{r}{9.47} &
    \multicolumn{1}{r}{3.32} &
    \multicolumn{1}{r}{2.05} &
    \multicolumn{1}{r}{26.84} &
    \multicolumn{1}{r}{3.92}\\
    \multicolumn{1}{r}{} &
    \multicolumn{1}{l}{EWC~\cite{kirkpatrick_rascanu}} & 
    \multicolumn{1}{c}{IQ} & 
     \multicolumn{1}{r}{2.04} &
     \multicolumn{1}{r}{28.21} & 
     \multicolumn{1}{r}{4.04}& 
     \multicolumn{1}{r}{2.11}& 
     \multicolumn{1}{r}{30.62}& 
    \multicolumn{1}{r}{4.03} & 
    \multicolumn{1}{r}{2.26} &
    \multicolumn{1}{r}{26.82} &
    \multicolumn{1}{r}{3.51} &
    \multicolumn{1}{r}{0.88} &
    \multicolumn{1}{r}{6.13} &
    \multicolumn{1}{r}{3.17} &
    \multicolumn{1}{r}{1.82} &
    \multicolumn{1}{r}{22.95} &
    \multicolumn{1}{r}{3.69}\\
    \multicolumn{1}{r}{} &
    \multicolumn{1}{l}{MAS~\cite{aljundi2018memory}} & 
     \multicolumn{1}{c}{IQ} &
    \multicolumn{1}{r}{2.18} & 
     \multicolumn{1}{r}{25.94} &
     \multicolumn{1}{r}{7.15} & 
     \multicolumn{1}{r}{2.30}& 
     \multicolumn{1}{r}{32.06}& 
     \multicolumn{1}{r}{7.34}& 
    \multicolumn{1}{r}{2.56} & 
    \multicolumn{1}{r}{31.58} &
    \multicolumn{1}{r}{4.97} &
    \multicolumn{1}{r}{1.13} &
    \multicolumn{1}{r}{7.97} &
    \multicolumn{1}{r}{3.58} &
    \multicolumn{1}{r}{2.04} &
    \multicolumn{1}{r}{24.39} &
    \multicolumn{1}{r}{5.76}\\
    \multicolumn{1}{r}{} &
    \multicolumn{1}{l}{Physics-Informed (ours)} & 
    \multicolumn{1}{c}{4} & 
     \multicolumn{1}{r}{1.99} &
     \multicolumn{1}{r}{26.74} & 
     \multicolumn{1}{r}{4.41}& 
     \multicolumn{1}{r}{2.09}& 
     \multicolumn{1}{r}{27.29}& 
    \multicolumn{1}{r}{4.36} & 
    \multicolumn{1}{r}{2.18} &
    \multicolumn{1}{r}{23.79} &
    \multicolumn{1}{r}{3.69} &
    \multicolumn{1}{r}{0.87} &
    \multicolumn{1}{r}{5.98} &
    \multicolumn{1}{r}{3.18} &
    \multicolumn{1}{r}{\textbf{1.78}} &
    \multicolumn{1}{r}{\textbf{20.95}} &
    \multicolumn{1}{r}{\textbf{3.91}}\\
\end{tabular}
\vspace{-0.2cm}
\end{center}
\end{table*}

\section{Evaluation}
\label{label_evaluation}

\paragraph{Training Setup} We used the standard SGD optimizer with a multi-step learning rate schedule, an initial learning rate of $0.001$, and a batch size of $64$. The performance of all methods is summarized in Table~\ref{table_dil_results}, which reports results for sequential training across four scenarios. All models were first trained for 50 epochs on the single-wall scenario, which includes all spatial grid locations. For each subsequent scenario, training was restricted to grid locations where additional walls were introduced, and the models were trained for 20 epochs. To mitigate catastrophic forgetting, a replay buffer was employed by randomly selecting $N^t$ samples from previously observed domains; we evaluate $N^t \in \{100, 500, 1000\}$. 
We report the MSE in relative position $\mathcal{E}_{p}$ (in m), azimuth $\mathcal{E}_{a}$ (in $^\circ$), and elevation $\mathcal{E}_{e}$ (in $^\circ$).

\begin{table}[!t]
\begin{center}
\setlength{\tabcolsep}{1.6pt}
    \caption{Evaluation of XceptionTime on $\{1, 2, 3, 4\}$ walls reporting the average MSE. \textbf{Bold} indicates lowest error.}
    \vspace{-0.1cm}
    \label{table_cross_validation}
    \begin{tabular}{ p{0.25cm} | p{0.25cm} | p{0.25cm} | p{0.25cm} | p{0.25cm} | p{0.25cm} | p{0.25cm} | p{0.25cm} | p{0.25cm} | p{0.25cm} | p{0.25cm} | p{0.25cm} | p{0.25cm} }
        \multicolumn{1}{c|}{\textbf{Train}} & \multicolumn{3}{c|}{\textbf{1 wall}} & \multicolumn{3}{c|}{\textbf{2 walls}} & \multicolumn{3}{c|}{\textbf{3 walls}} & \multicolumn{3}{c}{\textbf{4 walls}} \\
        \multicolumn{1}{c|}{\textbf{Set}} & \multicolumn{1}{c}{$\mathcal{E}_{p}$} & \multicolumn{1}{c}{$\mathcal{E}_{a}$} & \multicolumn{1}{c|}{$\mathcal{E}_{e}$} & \multicolumn{1}{c}{$\mathcal{E}_{p}$} & \multicolumn{1}{c}{$\mathcal{E}_{a}$} & \multicolumn{1}{c|}{$\mathcal{E}_{e}$} & \multicolumn{1}{c}{$\mathcal{E}_{p}$} & \multicolumn{1}{c}{$\mathcal{E}_{a}$} & \multicolumn{1}{c|}{$\mathcal{E}_{e}$} & \multicolumn{1}{c}{$\mathcal{E}_{p}$} & \multicolumn{1}{c}{$\mathcal{E}_{a}$} & \multicolumn{1}{c}{$\mathcal{E}_{e}$} \\ \hline
        \multicolumn{1}{c|}{\textbf{1 Wall}} & \multicolumn{1}{r}{\textbf{0.88}} & \multicolumn{1}{r}{\textbf{7.71}} & \multicolumn{1}{r|}{\textbf{2.86}} & \multicolumn{1}{r}{1.59} & \multicolumn{1}{r}{24.39} & \multicolumn{1}{r|}{3.01} & \multicolumn{1}{r}{4.20} & \multicolumn{1}{r}{69.38} & \multicolumn{1}{r|}{3.74} & \multicolumn{1}{r}{5.89} & \multicolumn{1}{r}{80.93} & \multicolumn{1}{r}{4.64} \\    
        \multicolumn{1}{c|}{\textbf{2 Walls}} & \multicolumn{1}{r}{1.38} & \multicolumn{1}{r}{21.00} & \multicolumn{1}{r|}{3.10} & \multicolumn{1}{r}{\textbf{0.86}} & \multicolumn{1}{r}{\textbf{6.11}} & \multicolumn{1}{r|}{\textbf{2.99}} & \multicolumn{1}{r}{5.11} & \multicolumn{1}{r}{88.31} & \multicolumn{1}{r|}{4.33} & \multicolumn{1}{r}{5.16} & \multicolumn{1}{r}{83.99} & \multicolumn{1}{r}{7.53} \\
        \multicolumn{1}{c|}{\textbf{3 Walls}} & \multicolumn{1}{r}{4.03} & \multicolumn{1}{r}{88.67} & \multicolumn{1}{r|}{3.66} & \multicolumn{1}{r}{3.90} & \multicolumn{1}{r}{81.23} & \multicolumn{1}{r|}{3.68} & \multicolumn{1}{r}{\textbf{0.94}} & \multicolumn{1}{r}{\textbf{6.01}} & \multicolumn{1}{r|}{\textbf{3.17}} & \multicolumn{1}{r}{2.25} & \multicolumn{1}{r}{31.05} & \multicolumn{1}{r}{3.69} \\
        \multicolumn{1}{c|}{\textbf{4 Walls}} & \multicolumn{1}{r}{3.75} & \multicolumn{1}{r}{74.95} & \multicolumn{1}{r|}{4.39} & \multicolumn{1}{r}{3.64} & \multicolumn{1}{r}{67.77} & \multicolumn{1}{r|}{4.33} & \multicolumn{1}{r}{3.20} & \multicolumn{1}{r}{53.94} & \multicolumn{1}{r|}{3.75} & \multicolumn{1}{r}{\textbf{0.98}} & \multicolumn{1}{r}{\textbf{6.70}} & \multicolumn{1}{r}{\textbf{3.31}} \\
    \end{tabular}
    \vspace{-0.25cm}
\end{center}
\end{table}

\paragraph{Evaluation of the Domain Shift} To assess the effect of domain shift, we first evaluate models under cross-validation on seen and unseen environments. Table~\ref{table_cross_validation} reports the average MSE values and shows a substantial performance drop when testing is performed on a domain different from the training domain. For instance, a model trained on the 4 walls scenario achieves a low azimuth error on the same domain ($\mathcal{E}_a = 6.70^\circ$), but the error increases markedly to $74.95^\circ$, $67.77^\circ$, and $53.94^\circ$ when evaluated on the 1, 2, and 3 walls scenarios, respectively. These results highlight the model's limited generalization and strong domain dependence, thereby motivating the use of a DIL approach.

\begin{figure}[!t]
    \centering
    \vspace{-0.25cm}
    \includegraphics[trim=10 10 10 10, clip, width=1.0\linewidth]{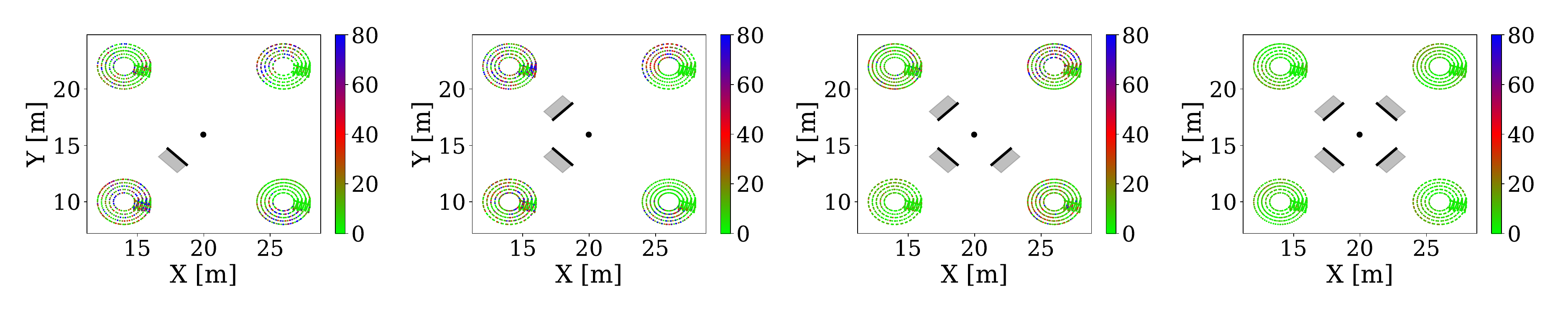}
    \vspace{-0.55cm}
    \caption{Evaluation of our physics-informed DIL method with an adaptation on 500 exemplars across $\{1,2,3,4\}$ walls.}
    \label{figure_evaluation_walls}
\end{figure}

\paragraph{Evaluation of DIL Techniques} After sequential training across all four domains, the model exhibits improved generalization (see Table~\ref{table_dil_results}). By leveraging physics knowledge as a domain-invariant feature for adaptation, it achieves better retention on earlier domains compared to Distillation, EWC, and MAS. For instance, with 100 exemplars it reaches $20.95^\circ$ azimuth and $3.91^\circ$ elevation, outperforming all baselines, and similarly achieves $14.95^\circ$ and $3.29^\circ$ with 500 exemplars and $13.14^\circ$ and $3.19^\circ$ with 1,000 exemplars. Figure~\ref{figure_evaluation_walls} shows the spatial distribution of azimuth estimation for our DIL model with physics-informed loss adaptation, indicating generally low errors over most trajectories but localized degradation near wall-obstructed regions as multipath becomes more severe.

\begin{figure}[!t]
    \centering
    \vspace{-0.4cm}
    \includegraphics[trim=10 10 10 10, clip, width=1.0\linewidth]{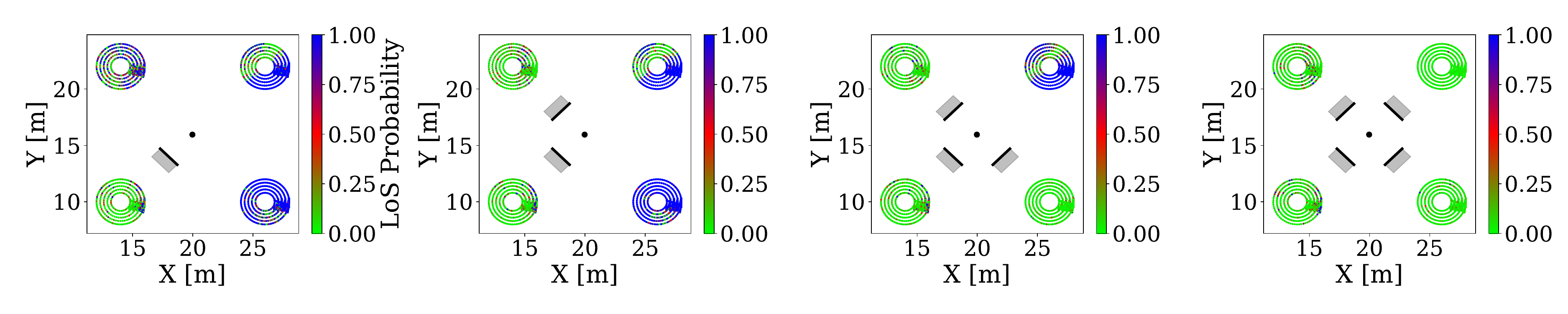}
    \vspace{-0.55cm}
    \caption{LoS/NLoS classification based on autoencoder-derived latent representations and unsupervised latent-space clustering.}
    \label{figure_los_nlos_classification}
    \vspace{-0.2cm}
\end{figure}

\begin{figure*}[!t]
    \centering
    \begin{minipage}[t]{0.243\linewidth}
        \centering
        \includegraphics[trim=10 9 4 10, clip, width=1.0\linewidth]{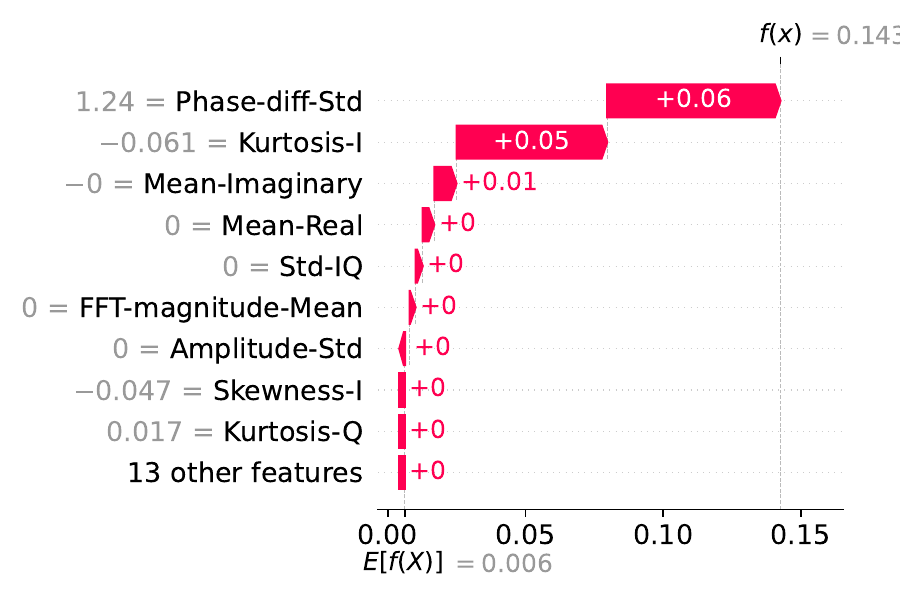}
        \vspace{-0.35cm}
        \caption{Feature analysis with SHAP on $n_f = 22$ features.}
        \label{figure_shapley_evaluation}
    \end{minipage}
    \hfill
    \begin{minipage}[t]{0.243\linewidth}
        \centering
        \includegraphics[trim=10 10 10 10, clip, width=1.0\linewidth]{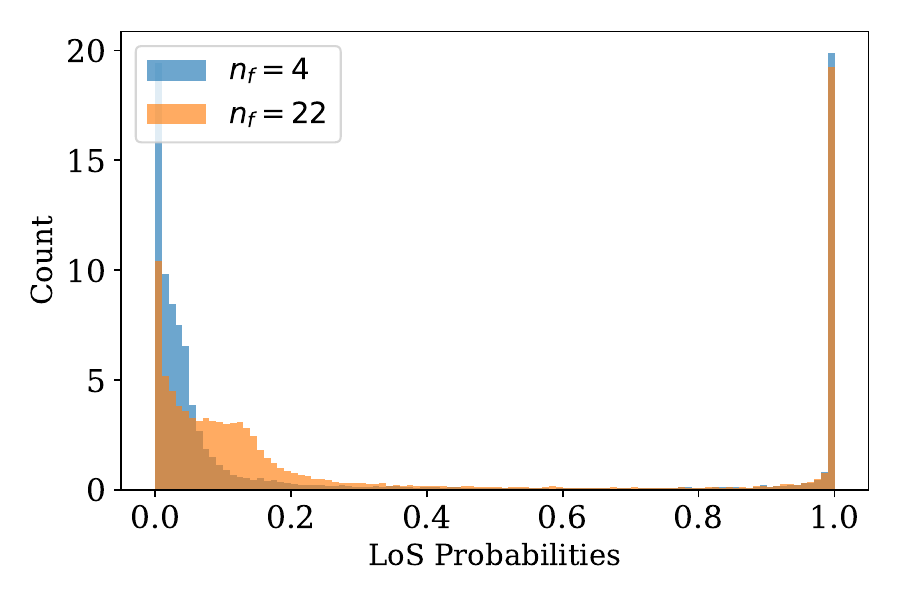}
        \vspace{-0.35cm}
        \caption{Histogram of LoS probabilities by the AE + GMM.}
        \label{figure_los_probability_walls}
    \end{minipage}
    \hfill
    \begin{minipage}[t]{0.243\linewidth}
        \centering
        \includegraphics[trim=10 12 10 12, clip, width=1.0\linewidth]{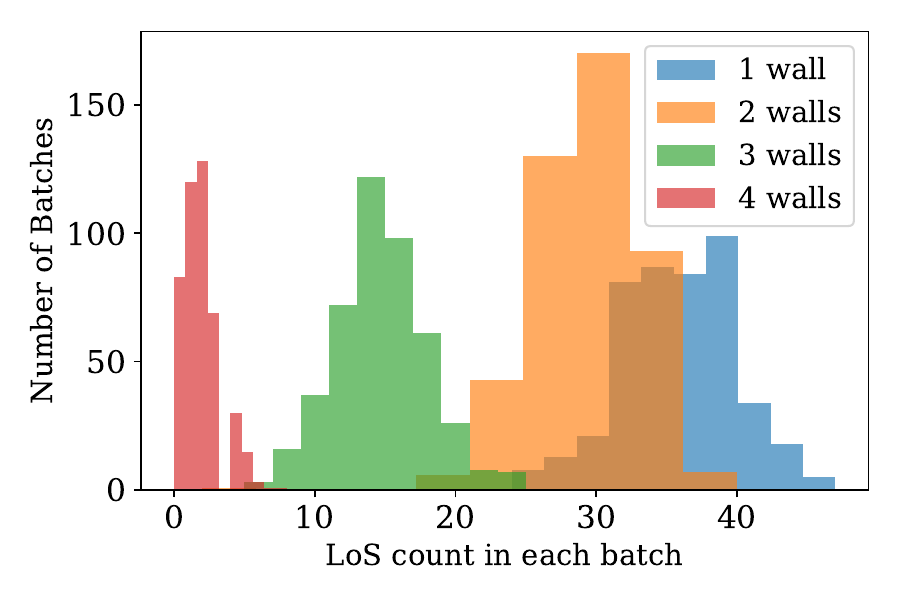}
        \vspace{-0.35cm}
    \caption{Number of LoS predictions per batch on each wall.}
    \label{figure_los_batch_count}
    \end{minipage}
    \hfill
    \begin{minipage}[t]{0.243\linewidth}
        \centering
        \includegraphics[trim=10 12 10 12, clip, width=1.0\linewidth]{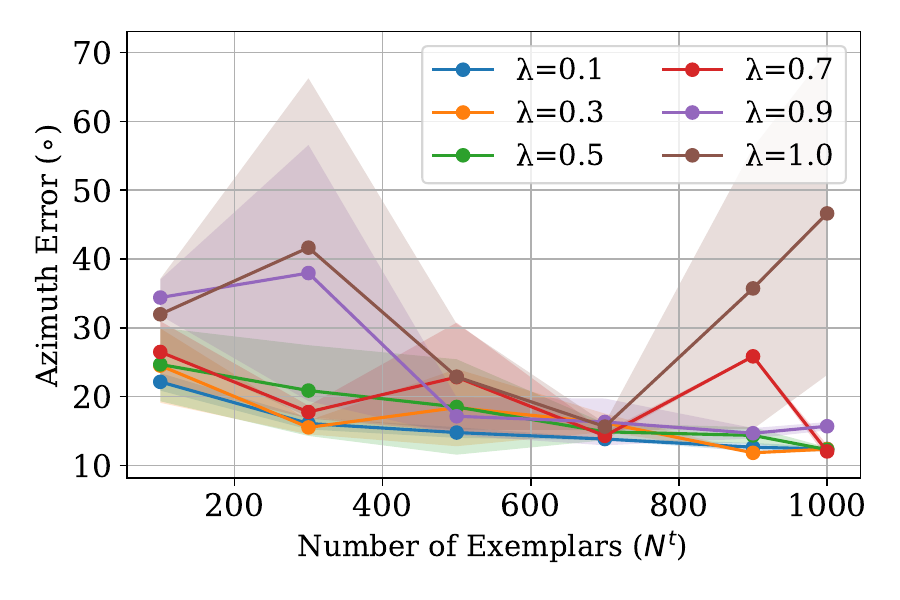}
        \vspace{-0.35cm}
    \caption{Parameter searches for $\lambda$ and exemplars number $N^t$.}
    \label{figure_hyperparameter_search}
    \end{minipage}
    \vspace{-0.25cm}
\end{figure*}

\paragraph{Evaluation of LoS/NLoS Classification} The proposed unsupervised LoS/NLoS classification is evaluated using the full feature set and a SHAP-reduced model. The AE and the subsequent Gaussian mixture model ($\mathrm{GMM}_{K=2}$) are trained exclusively on the training split of the two walls scenario. An AE is first trained for 20 epochs on all $88$ input features, corresponding to $n_f=22$ statistical features from each of the $M$ antenna elements, to learn a latent representation $\mathbf{z} \in \mathbb{R}^{4}$.  This latent space is then used as input to a GMM, to estimate posterior probabilities $\mathbf{p}$ for LoS/NLoS classification. As shown in Figure~\ref{figure_los_nlos_classification}, the framework effectively separates LoS and NLoS samples: regions without walls are assigned high LoS probabilities, whereas wall-blocked regions are classified as NLoS. For example, with one wall, the lower-left grid, which is separated from the jammer by a wall, yields the lowest LoS probability, while the remaining grids exceed $0.8$. With 4 walls, all samples are classified as NLoS with probabilities below $0.2$. These results show that the latent space of the AE combined with GMM clustering captures the spatial and environmental structure required for unsupervised LoS/NLoS classification.

\paragraph{Evaluation of SHAP} SHAP analysis of the composite function in Eq.~\ref{eq_kernelshap} yields attribution values $\boldsymbol{\psi} \in \mathbb{R}^{88}$ and shows that only $n_f=4$ features consistently dominate the LoS prediction (Figure~\ref{figure_shapley_evaluation}). Based on this finding, we construct a reduced AE using four features per antenna, resulting in a 16-dimensional input with the same latent dimension. The corresponding latent representation is again used as input to $\mathrm{GMM}_{K=2}$ for LoS/NLoS classification. As shown in Figure~\ref{figure_los_probability_walls}, both the full ($n_f{=}22$) and reduced ($n_f{=}4$) models assign consistently high LoS probabilities to LoS samples, while NLoS samples remain sufficiently separable despite slightly lower dispersion in the reduced model. Since the main objective is to identify LoS samples for constructing the mask $M_{\text{LoS}}$ in $\mathcal{L}_{p}$, this performance is sufficient for the proposed framework. The reduced model also preserves downstream localization accuracy, with errors changing only marginally from [$1.396\,\text{m}$, $13.141^\circ$, and $3.194^\circ$] to [$1.399\,\text{m}$, $13.499^\circ$, and $3.254^\circ$] for distance, azimuth, and elevation, respectively. These results show that the reduced feature set substantially lowers model complexity and computational cost while retaining the key information required for reliable LoS/NLoS classification and accurate jammer localization.

\begin{figure}[!t]
    \centering
    \begin{minipage}[t]{0.492\linewidth}
        \centering
        \includegraphics[trim=14 22 18 16, clip, width=1.0\linewidth]{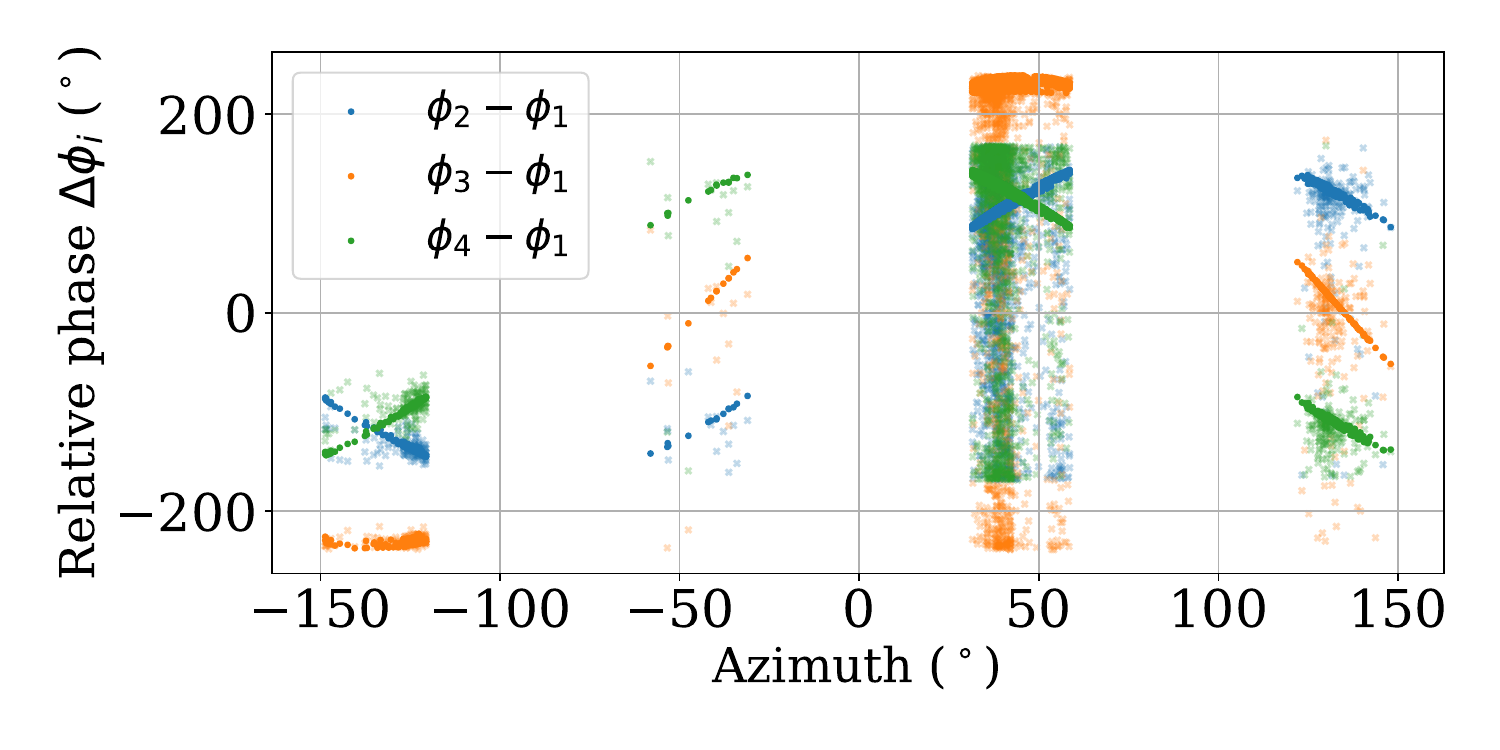}
        \label{figure_loc_phase_loss3}
    \end{minipage}
    \hfill
    \begin{minipage}[t]{0.492\linewidth}
        \centering
        \includegraphics[trim=14 22 18 16, clip, width=1.0\linewidth]{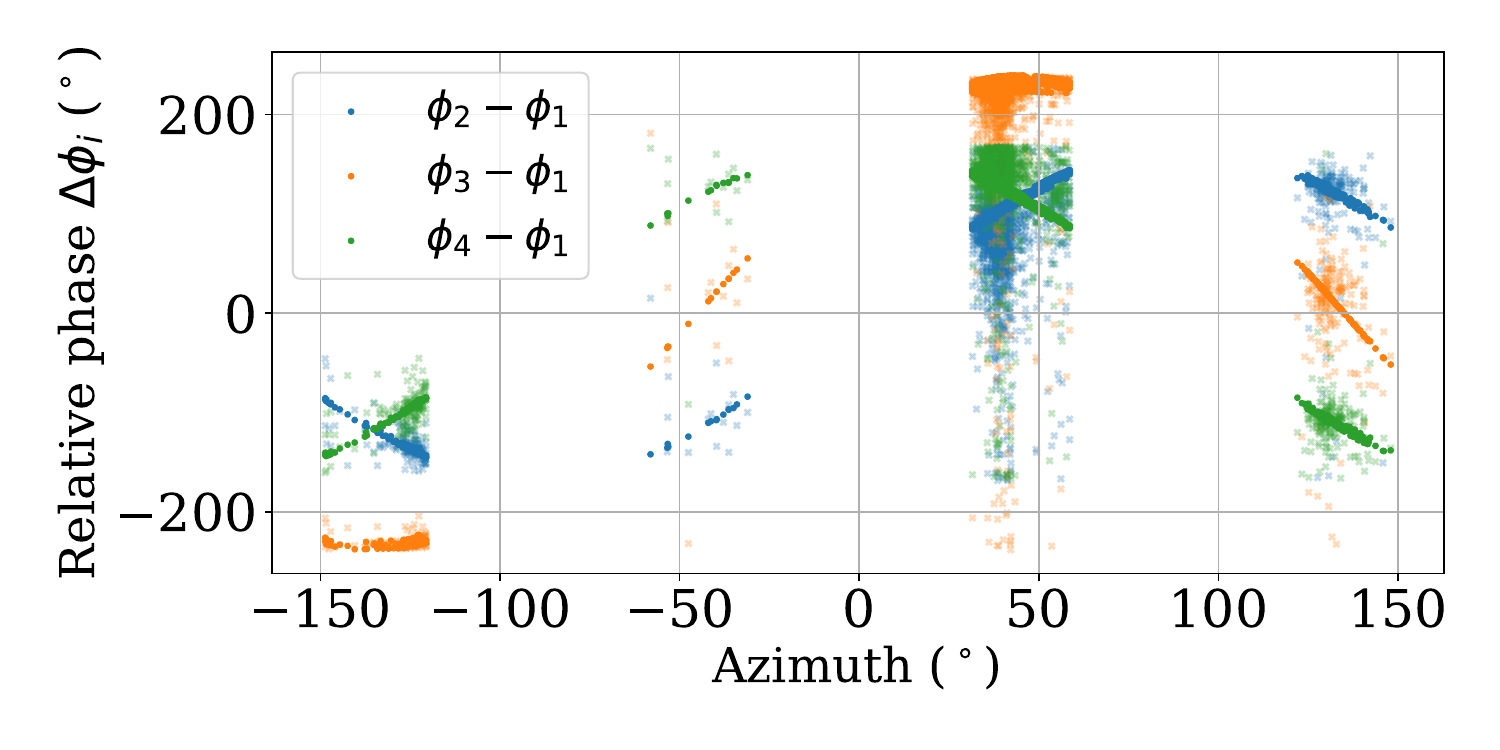}
        \label{figure_loc_phase_loss6}
    \end{minipage}
    \vspace{-0.4cm}
    \caption{Comparison of phase differences between antennas using $\mathcal{L}_l$ (left) and $\mathcal{L}_l + \lambda \mathcal{L}_p$ (right).}
    \label{figure_loc_phase_loss}
    \vspace{-0.1cm}
\end{figure}

\paragraph{Evaluation of the Physics-Informed Loss} The proposed physics-informed loss $\mathcal{L}_p$ regularizes the primary localization loss $\mathcal{L}_l$ by enforcing consistency of inter-antenna phase differences for LoS signals. Since this constraint is meaningful only under LoS propagation, $\mathcal{L}_p$ is applied exclusively to LoS samples ($\tau>0.9$) identified within each batch using a pre-trained $\mathrm{GMM}_{K=2}$. In practice, $\mathcal{L}_p$ occasionally vanished, particularly in the four-walls scenario, where batches often contained few or no LoS samples (Figure~\ref{figure_los_batch_count}), leading to unstable training. To address this, we enforced a minimum of five LoS samples from the replay buffer per batch during training on the four-wall scenario, which improved stability and performance. In the DIL setting, naive replay with randomly sampled past data yields limited generalization, whereas the proposed physics-informed regularization provides additional structural guidance across domains. Quantitatively, it reduces the average angular error by approximately $6^\circ$, $5^\circ$, and $3^\circ$ for replay buffers of $N= [100, 500, 1{,}000]$, respectively. The phase-plots in Figure~\ref{figure_loc_phase_loss} further show that models trained with $\mathcal{L}_p$ exhibit less scatter around the ground truth phase differences than models trained with $\mathcal{L}_l$ alone, indicating improved learning of physically consistent LoS phase relations.

\paragraph{Hyperparameter Searches} We study the effect of the physics-informed loss weight $\lambda$ in the DIL setting (average azimuth error over five random seeds), refer to Figure~\ref{figure_hyperparameter_search}. In low-exemplar regimes (100--300), performance is highly sensitive to $\lambda$: smaller values (0.1--0.3) yield lower errors, whereas larger values increase both error and variance, especially at $\lambda=1.0$. For larger exemplar sizes ($\geq 700$), performance becomes more stable and less sensitive to $\lambda$. Overall, $\lambda=0.3$ provides the best trade-off between stability and accuracy, particularly under limited memory.

\paragraph{Computation Times} EWC and MAS have the lowest iteration time ($0.08\,\text{s}$), but require $132\,\text{s}$ of per-domain overhead for parameter-importance estimation and $\approx24\,\text{MB}$ of memory for importance matrices. Distillation increases the iteration time to $0.15\,\text{s}$ and requires storing a previous model. In contrast, our physics-informed method is only slightly slower per iteration ($0.2\,\text{s}$), but uses a lightweight reusable $\text{f}_{\text{enc}}+\mathrm{GMM}_{K=2}$ pipeline ($33\,\text{KB}$ for $n_f=22$, $1.9\,\text{KB}$ for $n_f=4$, and $1.6\,\text{KB}$ for the GMM), introduces no post-training overhead, and avoids storing previous models, making it the most efficient overall.
\section{Conclusion}
\label{label_conclusion}

We presented a hybrid framework for GNSS interference AoA estimation that combines unsupervised LoS/NLoS classification, SHAP-guided feature reduction, and a selective physics-informed loss within a DIL setting. Real-world experiments across four wall-induced multipath domains showed improved cross-domain generalization, reducing angular error by up to $6^\circ$ in low-buffer regimes, retaining $17.63^\circ$ azimuth error after sequential adaptation, and preserving localization performance with only four selected features.
\blfootnote{\textbf{Acknowledgments.} This work has been carried out within the PaiL project, funding code 50NP2506, sponsored by the German Federal Ministry for Transport (BMV), and supported by the German Space Agency at DLR, the Bundesnetzagentur (BNetzA), and the Federal Agency for Cartography and Geodesy (BKG).}

\bibliography{ICL2026}
\bibliographystyle{IEEEtran}

\end{document}